\documentclass[preprint,unsortedaddress,showpacs,secnumarabic,amsmath,amssymb,prl]{revtex4-1}

\usepackage{graphicx}
\usepackage{color}

\begin{document}

\title{The interactions of same-row oxygen vacancies on rutile TiO$_2$(110)}

\author{B.~B. Kappes,$^{1}$ W.~B. Maddox,$^{1}$  D.~P. Acharya,$^2$ P. Sutter,$^2$ and C.~V. Ciobanu$^{1}$}
\thanks{Corresponding author, email: cciobanu@mines.edu}
\affiliation{$^1$Division of Engineering, Colorado School of Mines, Golden, CO  80401\\
$^2$Center for Functional Nanomaterials, Brookhaven National Laboratory, Upton, New York 11973}

\begin{abstract}

Based on a dipolar-elastic model for oxygen vacancies on rutile
(110), we evaluated analytically the overall energy of a periodic
array of two vacancies and extracted the interaction parameters from
total-energy density functional theory (DFT) calculations. Our
calculations show that the dipole model holds for next-nearest
neighbor vacancies and beyond. The elastic-dipolar interaction
vanishes for adjacent vacancies, but they still experience an
electrostatic repulsion. The proposed interaction model predicts a
vacancy separation distribution that agrees well with that
determined in our ultra-high vacuum scanning tunneling microscopy
experiments, and provides a perspective for understanding earlier
DFT reports.

\end{abstract}

\maketitle

Titanium dioxide --widely used in heterogeneous catalysis
\cite{heterog-catalysys}, photocatalysis \cite{photocatalysis},
solar cells \cite{solar}, or gas sensors \cite{sensors}, has become
the prototype material for studying the reactivity of metal oxide
surfaces \cite{prototype}. Defects such as oxygen vacancies are
always present on rutile surfaces \cite{vacpresent} and, depending
on their coverage and spatial distribution, can strongly influence
the reactivity of the surface \cite{bese-selloni}. The interactions
between vacancies determine their spatial distribution on the
surface. Highly reactive vacancy clusters or pairs have not been
expected to form because of vacancy repulsions \cite{pnnl-prl}, but
recent experiments \cite{cui-2008} do show the possibility of
spontaneously formed oxygen vacancy pairs (OVPs), {\em i.e.,} of two
adjacent vacancies in the same bridge-oxygen row. Regarding the
stability of OVPs, early density functional theory (DFT)
calculations came to contradictory conclusions. The OVPs were
reported to have the highest \cite{vijay-mills-metiu} and the lowest
\cite{rasmussen-molina-hammer} energy of all configurations of two
vacancies per computational cell. A newer study \cite{pnnl-prl}
finds virtually the same energies for the OVP and the next-nearest
neighbor (NNN) configurations, while another recent study
\cite{cui-2008} reports the NNN structure to have a much higher
energy than the OVP. To date, several issues have prevented the
complete, fundamental understanding of vacancy interactions,
including their reliable quantitative determination; the more
important issues are the difficulty of decoupling the interactions
while using computational slabs of manageable size, and the
sensitivity of various structural properties to the number of layers
in the supercells \cite{rasmussen-molina-hammer, OSC-vs-slabsize}.

Here we show that the interaction of same-row vacancies on rutile
(110) is dipolar-elastic in nature, with a long-range, inverse-cube
dependence on their separation. This dipolar-elastic model holds
when the vacancies are not adjacent, which we have found from DFT
calculations at the level of the generalized-gradient approximation
(GGA). Our approach has two key features that allow us to reliably
determine the formation energies and the interaction parameters from
total-energy GGA calculations: first, the interactions have been
isolated to one bridge-oxygen row by using large supercells, and
second, we have developed a closed-form expression for the overall
interaction (per computational cell) associated with a periodic
array of two vacancies. When vacancies are adjacent, they still
repel, but this repulsion is much weaker than the dipolar-elastic
one at the same distance. We have determined the distribution of
vacancy separations (along bridge-oxygen rows) by scanning tunneling
microscopy (STM), and have found that this distribution agrees well
with that predicted from the calculated interactions. This validates
our physical model for vacancy interactions, which we use to analyze
our DFT data as well as data from other works
\cite{cui-2008,pnnl-prl,rasmussen-molina-hammer}.

\begin{figure}
 \centering
\includegraphics[width=6.00cm]{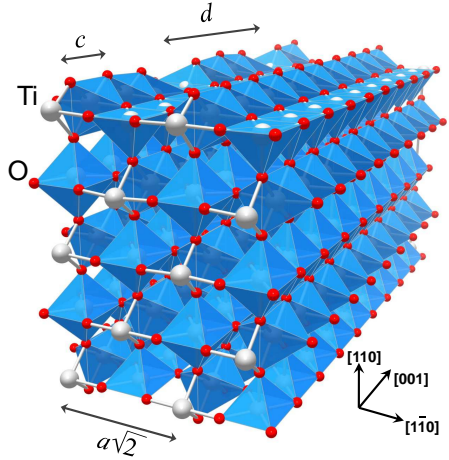}
\caption{(Color online) Reduced $10\times2$ rutile surface slab used
in the DFT calculations. The interaction between the two
bridge-oxygen vacancies is determined for different values of their
separation $d$, $c\leq d \leq 5c$.} \label{fig:geom}
\end{figure}

For the DFT simulations, we constructed $10\times 2$, 600-atom
stoichiometric supercells with dimensions $L_x=2a\sqrt{2}$ and
$L_y=10c$ (with $a=4.669$ \AA, $c=2.970$ \AA\
\cite{Morgan-Watson-rutilelattice2007}), and a thickness of five
O-Ti-O trilayers. The vacancies were created by removing two
same-row oxygen atoms spaced at $d$ ($c\leq d \leq 5c$)
[Fig.~\ref{fig:geom}]. The DFT relaxations were carried out in the
GGA framework using the PBE exchange-correlation functional
\cite{PBE}, projector-augmented wave \cite{PAW} pseudopotentials
\cite{VASP}, and an on-site Hubbard term $U$ for the Ti 3$d$ states
\cite{Dudarev}. Charge neutrality ($Q=0 e$) was maintained for the
stoichiometric slabs, but for the reduced slabs we also considered
the positively-charged case ($Q=4e$, corresponding to the removal of
two O$^{2-}$ ions). We have not searched for the localized electron
configurations that optimize the total energy \cite{locelectron},
but simply relaxed the structures from the bulk truncated positions
and analyzed their final electronic structures. For our non-zero
Hubbard term values, we have found that localization occurs on
subsurface Ti atoms for all spacings $d>c$.

\begin{figure}
 \centering
\includegraphics[width=14.0cm]{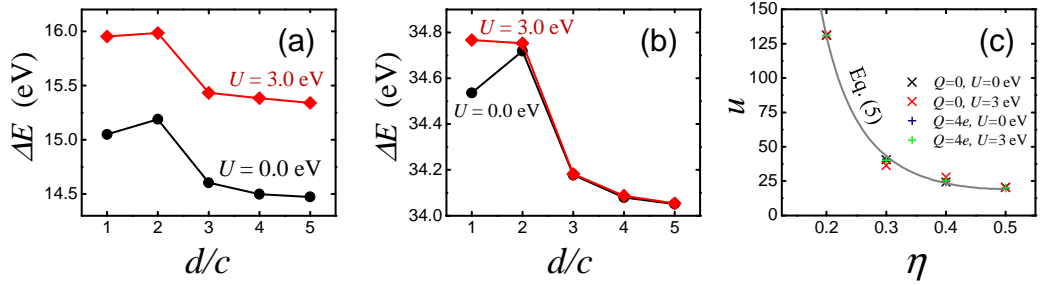}
\caption{(Color online) (a,b) The difference between the energy of a
reduced $(10\times 2)$ slab [(a) neutral, (b) positively charged]
and that of a stoichiometric one. (c) Analytical dependence
$u(\eta)$ (Eq.~(\ref{eq:riemann}), curve) and numerical calculations
of $u$ for neutral ($\times$ symbols) and charged slabs ($+$
symbols).} \label{fig:DeltaE}
\end{figure}

The difference $\Delta E$ between the total energy of the reduced
slab ($E_r$) and the energy of the stoichiometric one of same area
and thickness ($E_s$) can be written as
\begin{equation}
\Delta E \equiv E_r-E_s=  2 (f - \mu_{\mathrm{O}} ) + w,
\label{eq:Delta_E}
\end{equation}
where $f$ denotes the formation energy of a single vacancy (on an
otherwise perfect and wide surface), $\mu_{\mathrm{O}}$ is the
oxygen chemical potential, and $w$ contains all interactions. Since
we have collected all interactions into a single term $w$, which
depends on the supercell dimensions and on the spacing between the
vacancies, the formation energy $f$ in Eq.~(\ref{eq:Delta_E})
depends neither on the spacing between vacancies nor on their
coverage. The variation of $\Delta E$ with the separation $d$ at
constant $L_y$ ($L_y=10c$) is plotted in Figs.~\ref{fig:DeltaE}(a,b)
for neutral and positively charged slabs. In order to extract
interaction parameters from $\Delta E$ vs. $d$ data, we have to
understand the overall interaction term $w$.

Therefore, we first focus on finding an analytic expression for $w$,
and start by neglecting the cross-row interactions; this is
reasonable given the large supercell dimension along $[1\overline{1}
0]$, $L_x=2a\sqrt{2}=13.205$\AA. In this approximation, $w$
[Eq.~(\ref{eq:Delta_E})] depends only on the vacancy separation $d$
and on the dimension $L_y$ along a bridge-oxygen row. In the
framework of elasticity theory, point defects on surfaces interact
as elastic multipoles whose long-range interactions are inversely
proportional to certain powers of their separation
\cite{elasticmodels}. In what follows, we describe the interaction
$v$ between two {\em isolated} vacancies by the long-range
dipolar-elastic repulsion
\begin{equation} \label{eq:v_of_d}
v(d)  = \left\{ \begin{array}{rl}
 v_1 &\mbox{ if $d=c$} \\
 \frac{G}{d^3} &\mbox{ if $d=ic$, $i=2,3,4,...$}
       \end{array} \right. ,
\end{equation}
where $d$ is the distance between the two vacancies on an otherwise
perfect surface, $G$ is the strength of the dipolar repulsion, and
$v_1$ is a short-range interaction present only for adjacent
vacancies. When using periodic boundary conditions, the two
vacancies are not isolated, since they interact with their periodic
images as well. Using (\ref{eq:v_of_d}) for $d>c$ and collecting the
contributions from all periodic images along the same row, the total
interaction energy per supercell can be written as a function of
$L_y$ and $\eta\equiv d/L_y$ via
\begin{equation}
 w (L_y, \eta) = \frac{G}{L_y^3} u (\eta), \label{eq:w_and_u}
\end{equation}
with the function $u(\eta)$ given by
\begin{eqnarray}
 u(\eta) &   =    &  \frac{1}{\eta^3}+\sum^{\infty}_{k=1}\left(\frac{2}{k^3}+\frac{1}{(k+\eta)^3}+\frac{1}{(k-\eta)^3}\right)     \nonumber \\
         & \equiv & (1/\eta^3)+2\zeta(3)-(\psi^{(2)}(\eta)+\psi^{(2)}(-\eta))/2 \label{eq:u_series}
\end{eqnarray}
where $\zeta$ is the Riemann zeta function ($\zeta(3)\simeq 1.202$)
and $\psi^{(2)}$ is the polygamma function of second order.
Polygamma function identities \cite{reflection} reduce
Eq.~(\ref{eq:u_series}) to
\begin{equation}
u(\eta) =
2\zeta(3)-\pi^3\cot\left(\pi\eta\right)\csc^2\left(\pi\eta\right)-\psi^{(2)}\left(\eta\right).
\label{eq:riemann}
\end{equation}
Eq.~(\ref{eq:riemann}) is a general description of the interactions
of two identical, elastically-repelling defects in the same row and
their periodic images along that row; as such, it does not depend on
the (common) type of the defects ({\em e.g.}, both vacancies or both
adatoms), on their formation energy, or on their interaction
strength.

\begin{table}[thb]
\caption{Formation energies $f$ ($= \overline{f}+\mu_{\rm O}$),
repulsion strengths $G$, and short-range interactions $v_1$ for
different values of the slab charge $Q$ and Hubbard parameter $U$.
The standard deviations for $\overline{f}$ and $f$ are the same.}
\label{tab:fG}
\begin{ruledtabular}
\begin{tabular}{ccccc}
 $Q(e)$, $U$(eV) & $\overline{f}$(eV)  & $f$(eV)    & $G$(eV{\AA}$^3$) & $v_1$(eV) \\
\hline
 0, 0.0   & $ 7.170 \pm 0.006$ & 2.244 & $169.8 \pm 4.1$  & $0.677 \pm 0.013$ \\
 0, 3.0   & $ 7.611 \pm 0.013$ & 2.684 & $152.0 \pm 10.0$  & $0.702 \pm 0.027$ \\
 4, 0.0   & $16.966 \pm 0.004$ & 10.207 & $157.7 \pm 3.3$ & $0.575 \pm 0.009$ \\
 4, 3.0   & $16.964 \pm 0.006$ & 10.205 & $165.1 \pm 4.6$ & $0.808 \pm 0.013$ \\
\end{tabular}
\end{ruledtabular}
\end{table}

Using Eqs.~(\ref{eq:Delta_E}), (\ref{eq:w_and_u}), and
(\ref{eq:riemann}), we fit the data in Figs.~\ref{fig:DeltaE} (a,b)
for $d\ge 2c$ to obtain the relative formation energies
$\overline{f}\equiv f -\mu_{\mathrm{O}}$ and the interaction
strengths $G$ for different $Q$ (slab charge) and $U$ (Hubbard
parameter). The $\mu_{\rm O}$ values [see Eq.~(\ref{eq:Delta_E})]
that we have used were $\mu_{\rm O}=-4.926$ eV (half the energy of
an O$_2$ molecule) for the neutral system, and $\mu_{\rm O}=-6.759$
eV (the energy of an isolated O$^{2-}$ ion in the supercell) for
$Q=4e$. The calculated formation energies and interaction strengths
are listed in Table~\ref{tab:fG} for neutral and charged slabs at
$U=0.0$ eV and $U=3.0$ eV. Using $\overline{f}$ and $G$ values from
Table~\ref{tab:fG}, we check our model by numerically calculating
$u$ from Eq.~(\ref{eq:Delta_E}) [{\em i.e.}, $u=(\Delta E -
2{\overline f})L_y^3/G$ for $\eta \geq 2c/L_y$] and plotting it
along with the analytic result Eq.~(\ref{eq:riemann}). As seen in
Fig.~\ref{fig:DeltaE}(c), the agreement between the numerical $u$
values and the general formula Eq.~(\ref{eq:riemann}) is very good,
which validates {\em a posteriori} our assumption that the
interactions are reasonably well-confined to the bridge-oxygen row
as long as the neighboring rows are defect-free.

\begin{figure}
 \centering
\includegraphics[width=6.0cm]{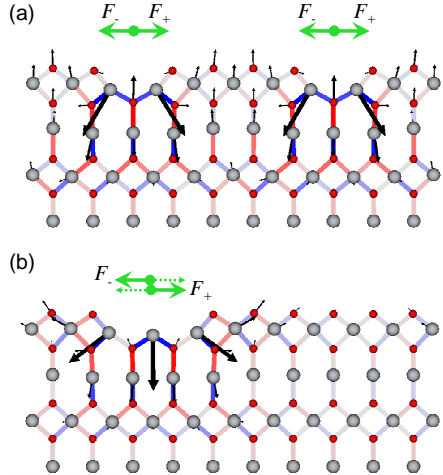}
\caption{(Color online)  Displacement fields in a plane containing
the vacancies separated by (a) $d=5c$ and (b) $d=c$; for clarity,
only three trilayers are shown. The green arrows show schematically
the horizontal force dipoles associated with each vacancy (a), and
illustrate the monopole cancelation responsible for the vanishing
elastic repulsion at $d=c$ (b). The displacements (magnified here
10-fold for clarity) are calculated for the ($Q=0e$, $U$=0eV) system
with respect to the relaxed stoichiometric slab. The largest
displacement magnitude is 0.41 \AA\ in (a) and 0.44\AA\ in (b).}
\label{fig:VV_VxV_VxxxxV}
\end{figure}

Although multipole interactions between atomic-level defects (most
often adatoms) on crystal surfaces have been studied
\cite{multipole}, so far the particular dipolar-elastic model
proposed here has not been reported for oxygen vacancies.
Figure~\ref{fig:VV_VxV_VxxxxV}(a) shows the atomic displacement
fields and, schematically, the horizontal force dipoles
($F_{+},F_{-}$) associated with each vacancy for $d=5c$. The atoms
located between vacancies experience opposite pulls resulting in an
increase of energy, {\em i.e.}, the elastic repulsion. When the
vacancies are brought close to form an OVP, there are no more 5-fold
coordinated Ti atoms (5-f Ti) between them, which leads to the
cancelation of two force monopoles as shown in
Fig.~\ref{fig:VV_VxV_VxxxxV}(b). It may be worth noting that
monopole cancelation has also been reported to be the origin of a
short-range attraction that leads to step bunching on certain
surfaces \cite{monopolecancellation}. Despite this monopole
cancelation, the interaction between vacancies at $d=c$ is not zero
but a quantity $v_1$ [Eq.~(\ref{eq:v_of_d})] that can be found from
a straightforward modification of Eq.~(\ref{eq:w_and_u}),
\begin{equation}
w(L_y, \eta_1\equiv \frac{c}{L_y}) = \frac{G}{L_y^3} u(\eta_1) -
\frac{G}{c^3} + v_1.   \label{eq:v1}
\end{equation}

Using Eqs.~(\ref{eq:Delta_E}), (\ref{eq:v1}) and the $\overline{f}$
and $G$ values already calculated, we have found that the
interaction $v_1$ is small but positive in all cases
[Table~\ref{tab:fG}, last column]. This short-range repulsion is
about one order of magnitude smaller than what the dipolar-elastic
model would predict for vacancies at $d=c$ [$G/c^3\approx 6.48$ eV],
and is largely due to the electrostatic interactions of the exposed
4-f Ti with the nearby 5-f Ti atoms.

\begin{figure}
 \centering
\includegraphics[width=7.5cm]{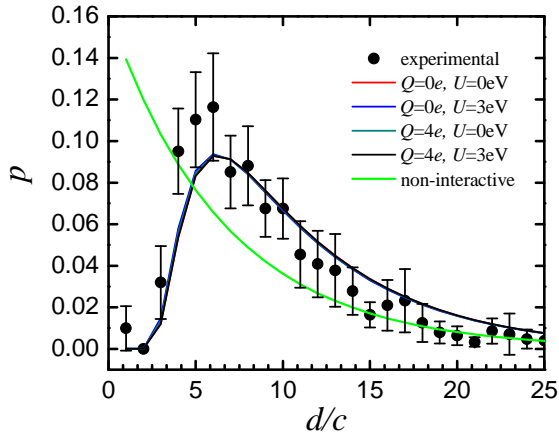}
\caption{(Color online) Distribution of vacancy spacings for
$n=15\%$ vacancy coverage at 950 K:  STM experiments (dots with
error bars) compared with the non-interacting system (exponential
decay, green curve) and with the canonical distribution (four nearly
overlapping curves). Note that the canonical distributions for
different $Q$ and $U$ values are not fits to the experimental data,
but are based on the interaction model Eq.~(\ref{eq:v_of_d}) with
the GGA values for $G$ and $v_1$ listed in Table~\ref{tab:fG}.}
\label{fig:stat}
\end{figure}

The directly observable manifestation of vacancy interactions is
their spatial distribution, which we have analyzed from thermal
vacancy populations. The samples were produced by Ar$^+$ sputtering
cycles, followed by annealing the (110) rutile surface at a
temperature $T=950$~K, then rapidly cooling down to 77 K to freeze
in the vacancy distribution. The vacancies were imaged using an
ultrahigh vacuum (UHV, pressure below 10$^{-11}$ Torr) cryogenic
STM, in constant current mode with positive sample bias
\cite{distinguish_from_OH}. Vacancy separations were analyzed from
portions of bridge-oxygen rows that had a vacancy concentration of
$n\approx 15\%$. If there were no interactions, then a fixed vacancy
coverage $n$ would lead to an exponential decay of the probability
to find vacancy-to-vacancy (V-V) segments of length $d$,
$p_{nonint}(d)\propto \exp{(-nd/c)}$ \cite{nonint-statistics}. Our
data shows that $p(d)$ exhibits a maximum, and not the monotonic
decay corresponding to the non-interacting system [green curve in
Fig.~\ref{fig:stat}]; this is a direct consequence of the repulsive
interactions between vacancies.

In a canonical ensemble system of V-V segments, the interactions
between the ends of segments give the single-particle energy levels
$v(d)$ ($d=c,2c,3c,...$) and thus a canonical distribution
$p(d)=(1/Z) \exp{(-nd/c)}\exp{(-v(d)/k_B T)}$, in which
$Z$ is a normalization factor and $k_B$ is Boltzmann's constant. The
canonical distribution based on the interaction model
Eq.~(\ref{eq:v_of_d}) with the parameters in Table~\ref{tab:fG} is
consistent with the experimental data [refer to
Fig.~\ref{fig:stat}], and is virtually the same for all ($Q$, $U$)
pairs used in this study. The competition between the fixed coverage
constraint and the rapidly-decreasing dipolar repulsion ($G/d^3$)
gives rise to a most-probable vacancy spacing $d^*$ which can be
readily derived from the canonical distribution, $d^* = (3Gc/nk_B
T)^{1/4}$. The $G$ values (Table~\ref{tab:fG}) give $d^*$ between
6.1$c$ and 6.3$c$, consistent with the experimental peak at 6$c$.
The agreement between the vacancy separation statistics determined
in STM and the canonical distribution with GGA-calculated parameters
validates our interaction model Eq.~(\ref{eq:v_of_d}).

In the previous systematic attempts to compute the vacancy
interactions on TiO$_2$(110)
\cite{vijay-mills-metiu,rasmussen-molina-hammer}, the total energy
was expressed as pairwise interactions from each vacancy to the next
one along the row and also included cross-row couplings. Both
reports \cite{vijay-mills-metiu,rasmussen-molina-hammer}
acknowledged unresolved shortcomings of the pairwise model, which
had manifested in significant differences of the total energies
predicted by the model (with respect to those obtained directly from
DFT calculations) once the model was applied to supercells other
than those used to determine the pairwise interaction parameters.
Departing from these pairwise models of
Refs.~\cite{vijay-mills-metiu,rasmussen-molina-hammer}, we have
proposed herein that the same-row vacancy interactions are dipolar,
thus long-ranged. As we will show below, our model
Eq.~(\ref{eq:w_and_u})--(\ref{eq:v1}) holds very well when applied
to different supercells (than those in Fig.~\ref{fig:geom}),
different numbers of vacancies (one or two) per cell, and different
exchange-correlation functionals. For example, we have used
Eqs.~(\ref{eq:w_and_u})--(\ref{eq:v1}) and the ($Q=0, U=0$) case
data in Table~\ref{tab:fG} to compute the total energy difference
between $5\times3$ supercells with two vacancies in the NNN and OVP
configurations. We have obtained 0.264 eV, in excellent agreement
with our GGA simulations performed for $5\times 3$ slabs, which
place the NNN supercell energy at 0.249 eV above that of the OVP
supercell; for the other three cases in Table~\ref{tab:fG}, the
model [Eqs.~(\ref{eq:w_and_u})--(\ref{eq:v1})] yields total energy
differences that are within 0.06 eV or less from the GGA results.
Our results for $5\times 3$ supercells are in quantitative
disagreement with the recently reported total energy difference of
0.8 eV \cite{cui-2008}, but based on our calculations and on the STM
experiments that show similarly small occurrence probability for
OVPs and NNNs (Fig.~\ref{fig:stat}), we believe the 0.8 eV value to
be in error.

Interestingly, data from other reports of DFT simulations
\cite{rasmussen-molina-hammer,pnnl-prl} can be readily understood
using our model. The two-vacancy results in Ref.~\cite{pnnl-prl}
($c<d<5c$) can be fitted well to our Eqs.~(\ref{eq:w_and_u}),
(\ref{eq:riemann}), yielding a strength $G= 244.9\pm 7.9$ eV\AA$^3$
(not enough data is provided to determine $f$ or $\overline f$). For
one vacancy per supercell, the quantity denoted as VFE in
\cite{rasmussen-molina-hammer} is defined as our $\Delta E \equiv
E_r - E_s$ up to an additive $\mu_{\mathrm O}$. We find that the
data for $p(m\times 1)$ cells ($m=2,3,4,6$)  in table 2 of
Ref.~\cite{rasmussen-molina-hammer} fits closely our dipolar-elastic
model, which for single-vacancy supercells takes the simple form
$\Delta E +\mu_{\mathrm O} = f +G \zeta(3)/L_y^3$. This data yields
$f = 3.134\pm 0.012$ eV and $G = 198.98 \pm 4.06 $ eV\AA$^3$; these
values are consistent with those in the first line of our
Table~\ref{tab:fG}, but differ from them likely because of the
different computational parameters used in
\cite{rasmussen-molina-hammer}. While we have devised our physical
model for interactions confined to the same row, the structures in
Refs.~\cite{pnnl-prl, rasmussen-molina-hammer} do not have any
intact oxygen row and thus allow for cross-row coupling. Even so,
the $p$($m\times 1$) data fits our elastic model very well, as
judged by the small standard deviations obtained for $f$ and $G$; in
the case of Ref.~\cite{rasmussen-molina-hammer} ($p(m\times 1)$
cells in table 2, $m\geq2$), this is because at one vacancy per
supercell, the cross-row interactions occur mostly perpendicular to
the oxygen rows and thus may amount to a constant independent of $m$
(the cell dimension along the row). For the two-vacancy results in
Ref.~\cite{pnnl-prl}, the agreement with our model likely occurs
because the diagonal cross-row interactions do not vary
significantly as a function of $d$ when $d>c$.

In conclusion, we have shown that the dipolar-elastic model
describes well the long-range repulsion of same-row vacancies for
all separations except $d=c$, where a much smaller short-range
interaction is present. The model fits not only our DFT data, but
also explains several other results from the literature and gives an
equilibrium vacancy separation distribution that agrees well with
that determined in our STM experiments.

{\em Acknowledgments.} Research carried out in part at the Center
for Functional Nanomaterials, Brookhaven National Laboratory, which
is supported by the U.S. Department of Energy, Office of Basic
Energy Sciences, under Contract No. DE-AC02-98CH10886. We
acknowledge funding from the DOE Office of Basic Energy Sciences,
Chemical Imaging Initiative FWP CO-023; funding from NSF through
Grants Nos. OCI-1048586 and CMMI-0846858; and access to
supercomputing resources at NCSA (Grant Nos. CHE-080019N and
DMR-090121) and at the Golden Energy Computing Organization.


\begin{thebibliography}{99}
\bibitem{nonint-statistics} The probability to encounter
$k$ vacancies in a row interval of length $d$ is given by the
Poisson distribution, $[(nd/c)^k\exp{(-nd/c)}]/k!$
\cite{nonintreference}. For $k=0$, this yields the probability to
find unintrerupted segments of length $d$,
$p_{nonint}(d)=(e^n-1)\exp{(-nd/c)}$, after normalizing such that
$\sum _{i=1}^{\infty} p(ic)=1$.
\end{thebibliography}
\end{document}